# Random bit generation based on self-chaotic microlasers with enhanced chaotic bandwidth

Jian-Cheng Li[1,2], Jin-Long Xiao[1,2], Yue-De Yang[1,2], You-Ling Chen[1,2], and Yong-Zhen Huang[1,2]*

[1] State Key Laboratory of Integrated Optoelectronics, Institute of Semiconductors, Chinese Academy of Sciences, Beijing 100083, China;

[2] Center of Material Science and Optoelectronic Technology, University of Chinese Academy of Sciences, Beijing 100049, China.

*Correspondence: Y. Z. Huang, E-mail: yzhuang@semi.ac.cn

**Abstract**: Chaotic semiconductor lasers have been widely investigated for high-speed random bit generation, which is applied for the generation of cryptographic keys for classical and quantum cryptography systems. Here, we propose and demonstrate a self-chaotic microlaser with enhanced chaotic bandwidth for high-speed random bit generation. By designing tri-mode interaction in a deformed square microcavity laser, we realize a self-chaotic laser caused by two-mode internal interaction, and achieve an enhanced chaotic standard bandwidth due to the photon-photon resonance effect by introducing the third mode. Moreover, 500 Gb/s random bit generation is realized and the randomness is verified by the NIST SP 800-22 statistics test. Our demonstration promises the applications of microlasers in secure communication, chaos radar, and optical reservoir computing, and also provides a platform for the investigations of multimode nonlinear laser dynamics.

## 1. Introduction

Physical random bits play an important role in cryptography systems, information security, stochastic modeling, and Monte Carlo simulation[1–5]. Physical random bit generation (RBG) was achieved with low generation rates (at Mb/s level) based on thermal noise[6] and sampling phase jitter[7] in specific circuits, stochastic threshold switching in memristors[8], and quantum vacuum state fluctuations[9]. To realize high-speed RBG, chaotic semiconductor lasers as favorable physical entropy sources have been widely investigated owing to their large bandwidth and intensive randomness[10–15]. However, semiconductor lasers, governed by the two parameters of mode intensity and carrier inversion, usually need external perturbations to generate specific nonlinear dynamic states, such as periodic oscillations and chaos[16]. Consequently, chaotic semiconductor lasers were investigated under external optical feedback[10,11,17–20] and optical injection[15,21,22]. To simplify the system complexity, integrated chaos lasers were developed with a passive feedback cavity[13,23–25], optoelectronic feedback[26], or mutual injection lasers[27,28]. In addition, deterministic polarization chaos, caused by nonlinear mode competition including carrier spin relaxation, was realized for a free-running quantum dot vertical-cavity surface-emitting laser[29]. Recently, parallel ultrafast RBG was demonstrated in a broad area semiconductor laser with curved facets, using spatiotemporal interference of many lasing modes with unpredictable spontaneous noise[30]. A self-chaotic microcavity laser was demonstrated by using two-mode internal interaction, and 10 Gb/s RBG was obtained from the chaotic laser output[31]. However, chaotic semiconductor lasers under delayed optical feedback or mutual coupling have obvious correlation peaks of the time delay signature[10,14], which reduces the randomness and security in random number generation. The ultrafast RBG relying on many modes requires a large broad area cavity under large-current pulse operation[30], and the chaotic bandwidth and RBG rate were limited by the laser relaxation oscillation frequency[31]. Based on the optical heterodyne for eliminating time-delay correlation, a 10 Gb/s real-time random bit generation was achieved with all-optical quantization[32]. To reduce correlation and enlarge bandwidth, complex chaotic signal system was demonstrated using continuous-wave laser and external-cavity laser with self-phase-modulated injection[33].

In this paper, we propose and demonstrate a tri-mode self-chaotic microlaser with an enhanced chaotic bandwidth by employing the photon-photon resonance effect[34]. By designing a deformed square microcavity with circular sides, we can enhance the mode $Q$-factors and engineer the mode frequency interval[35]. Since passive mode $Q$-factors are larger than $10^4$ for the fundamental ($0^{th}$), first ($1^{st}$) and second-order ($2^{nd}$) transverse modes, they can all approach the threshold condition for an AlGaInAs/InP deformed square microlaser, with a $Q$-factor determined by absorption loss much lower than $10^4$. As shown in Fig. 1(a), the self-chaotic microlaser is realized by the mode interaction between the $0^{th}$ and $1^{st}$ order transverse modes, and the chaotic bandwidth is enhanced due to photon-photon resonance caused by mode beating with the $2^{nd}$ order transverse mode. The mode intensity patterns of the $0^{th}$ and $1^{st}$ transverse modes are shown in the insets of Fig. 1(c), their field distributions are in-phase and anti-phase in half a region, respectively, which is clearer than those in the deformed hexagonal microcavity[31]. The enhancement and cancellation of mode beating intensities result in strong differences of carrier consumption in the in-phase and anti-phase regions, which transfer at the mode beating frequency. As the beating frequency approaches the relaxation oscillation frequency, the mode beating intensity will cause strong carrier oscillation and the appearance of mode side peaks similar to that under external modulation, which results in strong internal mode interaction and self-chaos[31]. The further mode beating with the $2^{nd}$ order transverse mode will induce additional high-frequency peaks in the response curve as shown in Fig. 1(a), i.e., the chaotic bandwidth enhanced by the photon-photon resonance effect for directly modulated lasers[34]. Based on the novel method, we demonstrate a tri-mode self-chaos deformed square microcavity laser with 33.9 GHz chaos bandwidth, and realize 500 Gb/s RBG from the chaotic microlaser output.

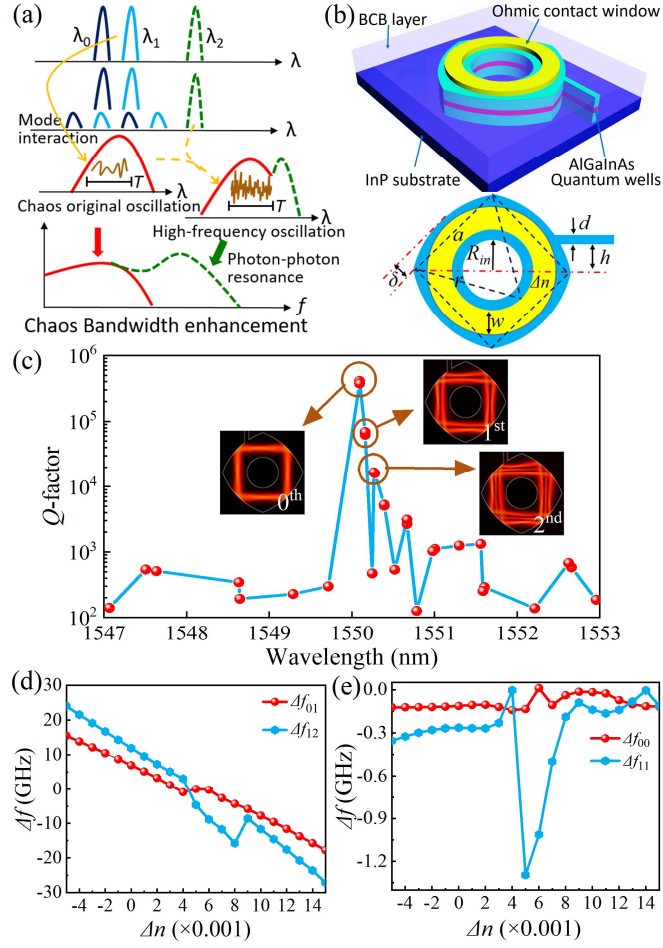

**Fig. 1** (a) Schematic diagram of self-chaos due to two-mode interaction and chaotic bandwidth enhanced by photon-photon resonance of the mode beating with the third mode. Carrier oscillation at the beating frequency of the $0^{th}$ and $1^{st}$ transverse modes $\lambda_0$ and $\lambda_1$ causes side peaks for the lasing modes, which work as the internal optical injection terms and cause self-chaos. The chaotic bandwidth is extended to the high-frequency region due to the photon-photon resonance with the $2^{nd}$ transverse mode $\lambda_2$. (b) Three-dimensional schematic diagram and two-dimensional top-view of a circular-sided square microcavity laser with a central hole and a ring electrode corresponding to a refractive index step $\Delta n$. (c) Mode $Q$-factor versus mode wavelength. Insets are mode intensity distributions of the $0^{th}$, $1^{st}$ and $2^{nd}$ transverse modes. (d) Transverse mode intervals $\Delta f_{01}$ and $\Delta f_{12}$ and (e) degenerated mode intervals $\Delta f_{00}$ and $\Delta f_{11}$ versus $\Delta n$.

## 2. Self-chaos generation

A three-dimensional schematic diagram and two-dimensional top-view of the deformed square microcavity laser are shown in Fig. 1(b), where a central hole is applied to further control the transverse mode number. The transverse electric (TE) mode characteristics are numerically investigated using a two-dimensional finite element method, for a deformed square with the flat-side length $a$ = 20 μm, circular-side deformation parameter $\delta$ = 2.17 μm, the width of output waveguide $d$ = 1.5 μm, the shift of the output waveguide $h = 4\sqrt{2}$ μm, and the radius of the central hole $R_{in}$ = 5.5 μm. Two degenerated modes with nearly the same magnitude of $Q$-factors and mode field patterns are obtained for each transverse mode. We give the results for the degenerated mode with a higher $Q$-factor in the following. As shown in Fig. 1(c), the simulated mode $Q$-factors are $4.1 \times 10^5$, $6.9 \times 10^4$, and $1.7 \times 10^4$ for the $0^{th}$, $1^{st}$ and $2^{nd}$ order transverse modes, respectively, with mode wavelengths of 1550.093, 1550.160, and 1550.265 nm. The corresponding squared magnetic field distributions are shown in the insets of Fig. 1(c). In addition, a ring p-electrode with a width of 4 μm is designed for fine adjustment of the mode frequency interval, with a refractive index step $\Delta n$ to simply account for carrier and temperature distributions inside the resonator[36]. The calculated mode frequency intervals $\Delta f_{01} = f_{0th} - f_{1st}$ and $\Delta f_{12} = f_{1st} - f_{2nd}$ and degenerate mode intervals $\Delta f_{00}$ and $\Delta f_{11}$ versus $\Delta n$ are plotted in Figs. 1(d) and 1(e), respectively. The magnitude of $\Delta f_{01}$ around 10 GHz is suitable for realizing a chaotic microlaser caused by internal mode interaction[29]. In the range $0.003 < \Delta n < 0.008$, complex mode coupling results in a large splitting for $\Delta f_{11}$.

According to the designed microcavity parameters, circular-sided square microcavity lasers were realized using an AlGaInAs/InP compressively-strained multiple quantum-well laser wafer with the same manufacturing process as in Ref. 31. The microlasers were tested at a heat sink temperature of 289 K using the experimental setup shown in Fig. 2(a). The output power coupled into a tapered single-mode fiber (SMF) and the applied voltage versus continuous-wave injection current are plotted in Fig. 2(b), where the insets are the scanning electron microscope image of an etched microcavity and the lasing spectra from 4 to 40 mA. A threshold current of 4 mA is estimated based on lasing spectra. Nonlinear dynamics of the laser output were investigated, including lasing spectra, radio-frequency (RF) spectra, and time domain signal. As shown in Fig. 2(c), three peaks at 1539.384, 1539.560 and 1539.876 nm are observed at an injection current of 5.6 mA, with mode frequency intervals of 22 and 39.5 GHz. Comparing the simulated results in Fig. 1(c), these peaks are identified as the $0^{th}$, $1^{st}$ and $2^{nd}$ transverse modes, respectively. The corresponding RF spectrum is shown in Fig. 2(d), which almost coincident with noise floor at 5.6 mA. By increasing the current to 6.6 mA, side peaks with an interval of ~0.04 nm (~5 GHz) are observed for the main lasing peaks, which may be attributed to the mode beating between the degenerate modes of the $1^{st}$ transverse mode as indicated by the simulated results in Fig. 1(e). A sharp harmonic peak at 5 GHz appears in the corresponding RF spectrum in Fig. 2(d) at 6.6 mA. At 8.8 mA, a broadened lasing spectrum appears due to strong mode interaction, similar as the chaotic lasing spectrum in Ref. 31, which is mainly caused by the mode interaction between the $0^{th}$ and $1^{st}$ modes. The chaotic standard bandwidth, which covers 80% of the total RF power[37], is calculated to be 9.6 GHz at 8.8 mA.

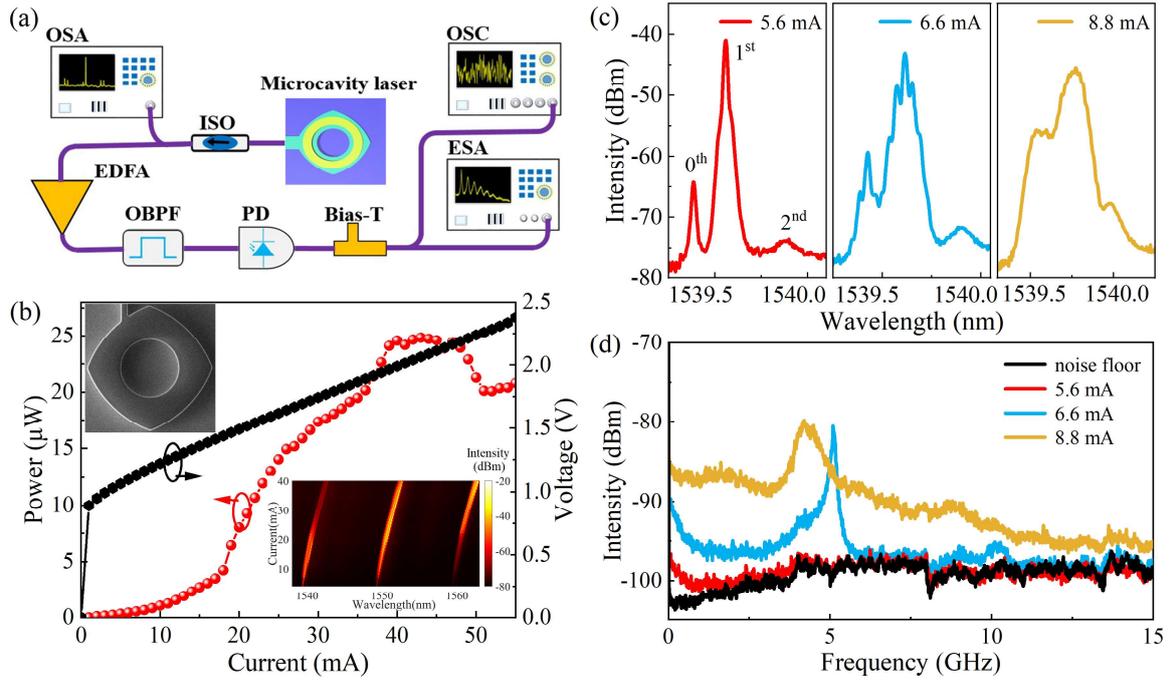

**Fig. 2** (a) Schematic of the experimental setup for the test of nonlinear dynamic states. ISO, isolator; OSA, optical spectrum analyzer; EDFA, erbium-doped fiber amplifier; OBPF, optical bandpass filter; PD, photodetector; ESA, electrical spectrum analyzer; OSC, real-time oscilloscope. (b) Laser power and applied voltage versus injected current. Insets: SEM image of a deformed square microcavity and lasing spectra map with respect to current. (c) Lasing spectra and (d) corresponding electric power spectra of steady, periodic, and chaotic states at 5.6, 6.6 and 8.8 mA, respectively.

## 3. Chaotic bandwidth enhancement

The enhancement of chaotic bandwidth due to photon-photon resonance is demonstrated in Fig. 3. Here, the main lasing modes jump to around 1550.5 nm with even high injection currents due to the current heating effect. As shown in Figs. 3(a) and 3(b), a long-wavelength mode assigned as the 2nd mode increases much faster than other lasing peaks with the current and becomes the main lasing mode at 20 mA, and the high frequency peaks at around 21 and 32 GHz of the RF spectra are greatly enhanced. The calculated chaos standard bandwidths are 13.7, 28.2, and 33.9 GHz at 16, 18, and 20 mA, respectively. To clearly verify the effect of photon-photon resonance, we measured RF spectra for filtered optical spectra as shown in Figs. 3(c) and 3(d). The RF spectra have small chaos standard bandwidths of 13.2 GHz and 8.1 GHz for the filtered optical spectra with the 0th plus 1st modes (0th + 1st) and 2nd mode, respectively. In Fig. 3(c), the intervals between the 2nd mode peak and three evident peaks of the wide chaotic spectra are 0.184, 0.272, and 0.320 nm, which contribute to three beating peaks at 22.8, 33.3, and 39.5 GHz for the RF spectrum in Fig. 3(d). These results imply the origin of bandwidth enhancement due to mode beating with the 2nd mode. The AC waveform of the chaotic laser output at 20 mA is plotted in Fig. 3(e), and the calculated autocorrelation function (ACF) is shown in Fig. 3(f), with a half width at half maximum of 0.011 ns. The ACF has some minor peaks within 0.5 ns, but without the time-delayed correlation peak observed in optical feedback chaotic lasers[38]. The modified Grassberger-Procaccia (G-P) algorithm is applied to quantify the complexity of the chaos signal[39,40], and a correlation dimension of 11.6 from Fig.4 is obtained, which is nearly triple that in Ref. 31.

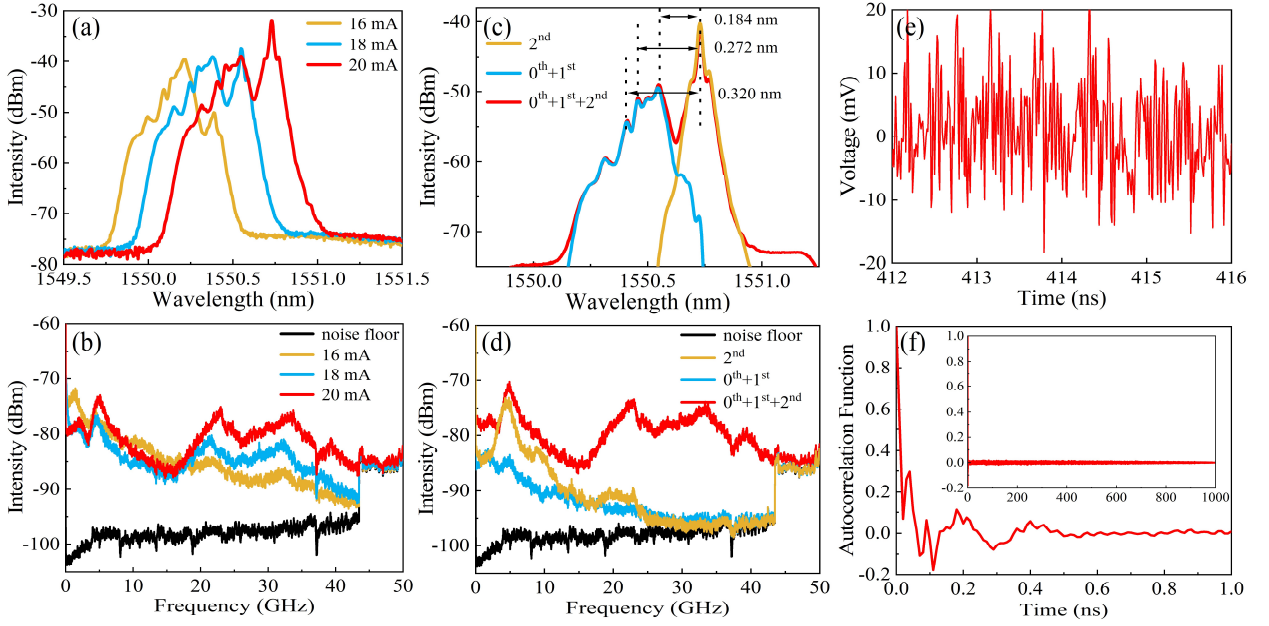

**Fig. 3** (a) Lasing spectra and (b) corresponding RF spectra at 16, 18 and 20 mA. (c) Filtered lasing spectra, the arrows show different peak intervals, and (d) corresponding RF spectra at 20 mA. (e) Irregular temporal waveform and (f) corresponding autocorrelation function for the chaotic output at 20 mA. The inset in (f) represents the entire ACF curve for 1 μs.

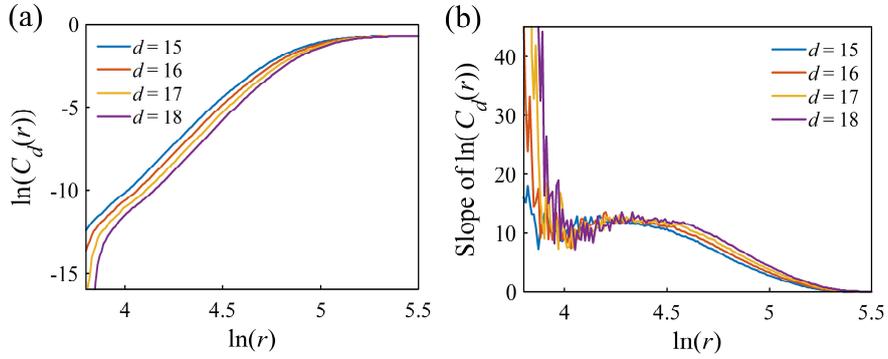

**Fig. 4** The correlation integral $C_d(r)$ versus the radius of the ball in logarithmic scale, and (b) corresponding slope of $\ln(C_d(r))$ curve.

## 4. Random bit generation

Furthermore, the bandwidth-enhanced self-chaotic microlaser was utilized to generate physical random numbers. The AC waveform signals at 20 mA are collected with a 100 GSa/s sampling rate, and the intensity histogram distribution of the 500 μs long raw data stream is illustrated in Fig. 5(a). The intensity distribution is asymmetric with an initial skewness of 0.40, which is a typical feature of a chaotic semiconductor laser. The asymmetric distribution can result in bias in the generated random sequence, and we adopted extra post-processing methods, including delay-subtracting and least significant bits (LSBs) extraction, for RBG[11,30]. Specifically, we subtract the original signal from its delayed signal to attain a symmetric distribution. Considering the very low correlation coefficient at 0.5 ns in Fig. 3(f), we select a delay time of 0.5 ns and plot the histogram distribution of the differential data in Fig. 5(b). The symmetry of the differential signal is significantly improved with a skewness coefficient of 0.02. Then, the differential intensity is digitalized into 8-bit binary numbers, and the LSBs method is adopted to destroy the residual correlations of adjacent bits and improve the uniformity of the bit distributions. By retaining five LSBs, we can obtain a nearly uniform probability distribution, as shown in Fig. 5(c). At the same time, the autocorrelation coefficient of the bit stream is less than $10^{-3}$ and remains at the background level for any bit stream length in Fig. 5(d), indicating the removal of correlation between successive bits.

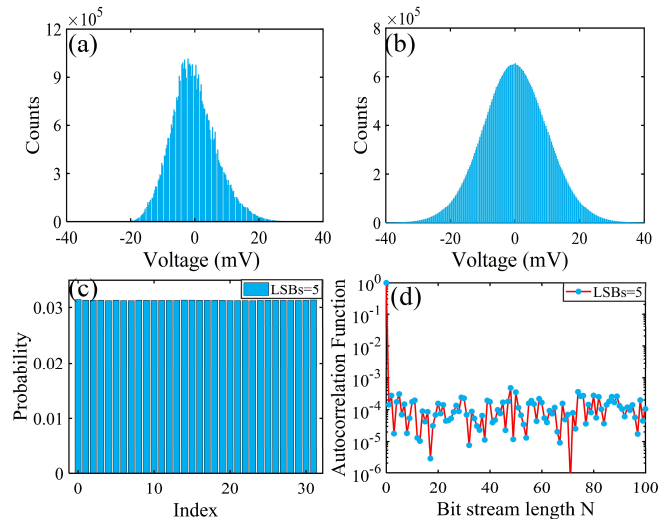

**Fig. 5** Histogram distribution for (a) raw signal intensity and (b) differential intensity after delay-subtracting post-processing. (c) Probability distribution with 5-LSBs extraction, and (d) corresponding ACF curve of the bit stream.

Finally, the randomness of the generated random bits is verified using the NIST Special Publication 800-22 statistical tests, by dividing 1-Gbit data into 1000 sequences of 1-Mbit[41]. When the significance level is set to 0.01, the randomness test is successful if the P-value is larger than 0.0001 and the proportion is within $0.99 \pm 0.0094392$. For the test items that produce multiple P-values and proportions, the worst case is selected and shown in Table 1, and the generated random bits successfully pass the 15 NIST sub-tests. The obtained maximum electrical-delay self-difference RBG rate is 500 Gb/s (100 GSa/s×5 bits).

**Table 1** NIST SP 800-22 test results for random bit sequence

| Statistical Test | Electrical-delay self-difference RBG | | Optical-delay self-difference RBG | | Result |
| --- | --- | --- | --- | --- | --- |
| | P-value | Proportion | P-value | Proportion | |
| Frequency | 0.32214 | 0.985 | 0.60799 | 0.992 | Success |
| Block Frequency | 0.27027 | 0.989 | 0.80556 | 0.986 | Success |
| Runs | 0.37701 | 0.987 | 0.16080 | 0.989 | Success |
| Longest Run | 0.29109 | 0.989 | 0.14532 | 0.986 | Success |
| Rank | 0.62255 | 0.983 | 0.44655 | 0.992 | Success |
| FFT | 0.40296 | 0.987 | 0.14781 | 0.988 | Success |
| Nonoverlaping template | 0.00798 | 0.989 | 0.00487 | 0.986 | Success |
| Overlapping template | 0.30266 | 0.986 | 0.30412 | 0.990 | Success |
| Universal | 0.60177 | 0.987 | 0.11606 | 0.990 | Success |
| Linear complexity | 0.23927 | 0.989 | 0.14125 | 0.986 | Success |
| Serial | 0.13650 | 0.990 | 0.13572 | 0.991 | Success |
| Approximate entropy | 0.01395 | 0.993 | 0.99743 | 0.990 | Success |
| Cumulative sums | 0.04365 | 0.986 | 0.14206 | 0.987 | Success |
| Random excursions | 0.12120 | 0.986 | 0.01572 | 0.987 | Success |
| Random excursions variant | 0.05205 | 0.986 | 0.00714 | 0.995 | Success |

We also conducted an optical-delay self-difference experiment for random bit generation via balanced-detection method in Ref. 42. As shown in Fig. 6, the chaotic light from the microcavity laser at 20 mA is firstly amplified and filtered. Then, the light is split into two paths after a 50:50 fiber coupler (FC). Delayed fiber (DL) with the length of 1 m (corresponding to 5 ns optical delay) is introduced into one of the two paths. The two beams are simultaneously detected by a balanced detector (Finisar BPD V2120R, 43 GHz bandwidth). Then the converted electrical signal is collected by the real-time oscilloscope at 100 GSa/s sampling rate. Finally, a least-significant-bits method is adopted and 5-LSBs are selected to generate 500 Gb/s physical random number sequence. Similarly, 1000 sequences of 1-Mbit stream are set to the NIST SP 800-22 randomness test. All sub-tests are successful and shown in Table 1.

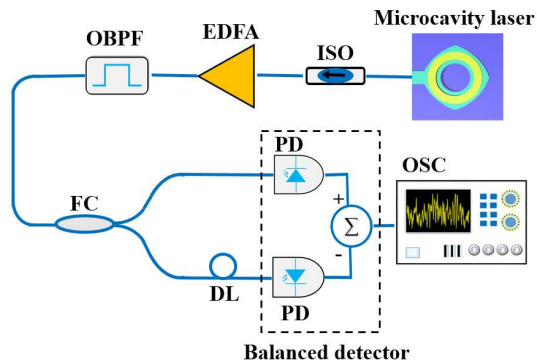

**Fig. 6** Experimental setup for optical-delay self-difference random bit generator. FC, fiber coupler; DL, delayed fiber.

## 5. Conslusion

In summary, tri-transverse-mode lasing with mode intervals of around 10 to 30 GHz has been demonstrated in a deformed square microcavity laser. The strong mode interaction between the $0^{th}$ and $1^{st}$ order transverse modes with the oscillation of the local photon density distribution inside the microcavity results in self-chaotic laser output, as in Ref. 31. Moreover, the $2^{nd}$ order transverse mode can induce additional beating peaks for the chaotic RF spectrum and greatly enhance the chaotic signal bandwidth. Based on tri-mode chaotic output, we have realized 500 Gb/s physical random number generation using electrical and optical delay-subtracting RBG schemes. Our work paves the way for mode engineering to enhance the self-chaotic bandwidth for deformed microcavity lasers. A random number generator based on the self-chaos deformed square laser can simplify the system greatly due to a small footprint and low power consumption for the chaotic microlasers. Moreover, self-chaotic lasers have potential applications in secure communication, chaos radar and optical reservoir computing.

## References

1. N. Gisin, G. Ribordy, W. Tittel, and H. Zbinden, "Quantum cryptography," *Rev. Mod. Phys.* **74**, 145-195 (2002)
2. S. Asmussen and P.W. Glynn, *Stochastic Simulation: Algorithms and Analysis*, Springer, New York, (2007)
3. M. Herrero-Collantes and J. C. Garcia-Escartin, "Quantum random number generators," *Rev. Mod. Phys.* **89**, 015004 (2017)
4. Y. Liu, X. Yuan, M.-H. Li, W. Zhang, Q. Zhao, J. Zhong, Y. Cao, Y. H. Li, L. K. Chen, H. Li, T. Peng, Y.-A. Chen, C. Z. Peng, S. C. Shi, Z. Wang, L. You, X. Ma, J. Fan, Q. Zhang, and J. W. Pan, "High-speed device-independent quantum random number generation without a detection loophole," *Phys. Rev. Lett.* **120**, 010503 (2018)
5. H. Gao, A. B. Wang, L. S. Wang, Z. W. Jia, Y. Y. Guo, Z. S. Gao, L. S. Yan, Y. W. Qin, and Y. C. Wang, "0.75 Gbit/s high-speed classical key distribution with mode-shift keying chaos synchronization of Fabry-Perot lasers," *Light-Sci. Appl.* **10**, 172 (2021).
6. C. S. Petrie and J. A. Connelly, "A noise-based ic random number generator for applications in cryptograph generator for applications in cryptography," *IEEE Trans. Circuits Syst. I-Fundam. Theor. Appl.* **47**, 615-621 (2000)
7. B. Sunar, W. J. Martin, and D. R. Stinson, "A provably secure true random number generator with built-in tolerance to active attacks," *IEEE Trans. Comput.* **56**, 109-119 (2007)
8. G. M. Kim, J. H. In, Y. S. Kim, H. Rhee, W. Park, H. C. Song, J. Park, and K. M. Kim, "Self-clocking fast and variation tolerant true random number generator based on a stochastic mott memristor," *Nat. Commun.* **12**, 2906 (2021)
9. C. Gabriel, C. Wittmann, D. Sych, R. Dong, W. Mauerer, U. L. Andersen, C. Marquardt, and G. Leuchs, "A generator for unique quantum random numbers based on vacuum states," *Nat. Photonics* **4**, 711-715 (2010)
10. A. Uchida, K. Amano, M. Inoue, K. Hirano, S. Naito, H. Someya, I. Oowada, T. Kurashige, M. Shiki, S. Yoshimori, K. Yoshimura, and P. Davis, "Fast physical random bit generation with chaotic semiconductor lasers," *Nat. Photonics* **2**, 728-732 (2008)
11. I. Reidler, Y. Aviad, M. Rosenbluh, and I. Kanter, "Ultrahigh-Speed Random Number Generation Based on a Chaotic Semiconductor Laser," *Phys. Rev. Lett.* **103**, 024102 (2009)
12. I. Kanter, Y. Aviad, I. Reidler, E. Cohen, and M. Rosenbluh, "An optical ultrafast random bit generator, " *Nat. Photonics* **4**, 58-61 (2010)
13. G. Verschaffelt, M. Khoder, and G. Van der Sande, "Random number generator based on an integrated laser with on-chip optical feedback," *Chaos* **27**, 114310 (2017)
14. S. Y. Xiang, B. Wang, Y. Wang, Y. A. Han, A. J. Wen, and Y. Hao, "2.24-Tb/s Physical Random Bit Generation with Minimal Post-Processing Based on Chaotic Semiconductor Lasers Network," *J. Lightwave Technol.* **37**, 3987-3993 (2019)
15. X. Z. Li and S. C. Chan, "Random bit generation using an optically injected semiconductor laser in chaos with oversampling," *Opt. Lett.* **37**, 2163-2165 (2012)
16. M. Sciamanna and K. A. Shore, "Physics and applications of laser diode chaos," *Nat. Photonics* **9**, 151-162 (2015)
17. T. Mukai and K. Otsuka, "New route to optical chaos: Successive-subharmonic-oscillation cascade in a semiconductor laser coupled to an external cavity," *Phys. Rev. Lett.* **55**, 1711-1714 (1985)
18. J. Mork, J. Mark, and B. Tromborg, "Route to chaos and competition between relaxation oscillations for a semiconductor laser with optical feedback," *Phys. Rev. Lett.* **65**, 1999-2002 (1990)
19. L. Jumpertz, K. Schires, M. Carras, M. Sciamanna, and F. Grillot, "Chaotic light at mid-infrared wavelength," *Light-Sci. Appl.* **5**, e16088 (2016).
20. Y. Deng, Z. F. Fan, B. B. Zhao, X. G. Wang, S. Zhao, J. Wu, F. Grillot, and C. Wang, "Mid-infrared hyperchaos of interband cascade lasers," *Light-Sci. Appl.* **11**, 7 (2022)

21. T. B. Simpson, J. M. Liu, A. Gavrielides, V. V. Kovanis, and P. M. Alsing, "Period-doubling cascades and chaos in a semiconductor laser with optical injection," *Phys. Rev. A* **51**, 4181-4185 (1995)
22. F. Y. Lin, S. Y. Tu, C. C. Huang, and S. M. Chang, "Nonlinear dynamics of semiconductor lasers under repetitive optical pulse injection," *IEEE J. Sel. Top. Quantum Electron* **15**, 604-611 (2009)
23. A. Argyris, M. Hamacher, K. E. Chlouverakis, A. Bogris, and D. Syvridis, "Photonic integrated device for chaos applications in communications," *Phys. Rev. Lett.* **100**, 194101 (2008)
24. J. G. Wu, L. J. Zhao, Z. M. Wu, D. Lu, X. Tang, Z. Q. Zhong, and G. Q. Xia, "Direct generation of broadband chaos by a monolithic integrated semiconductor laser chip," *Opt. Express* **21**, 23358-23364 (2013)
25. S. Sunada, T. Fukushima, S. Shinohara, T. Harayama, K. Arai, and M. Adachi, "A compact chaotic laser device with a two-dimensional external cavity structure," *Appl. Phys. Lett.* **104**, 241105 (2014)
26. P. Munnelly, B. Lingnau, M. M. Karow, T. Heindel, M. Kamp, S. Höfling, K. Lüdge, C. Schneider, and S. Reitzenstein, "On-chip optoelectronic feedback in a micropillar laser-detector assembly," *Optica* **4**, 303-306 (2017)
27. L. X. Zou, B. W. Liu, X. M. Lv, Y. D. Yang, J. L. Xiao, and Y. Z. Huang, "Integrated semiconductor twin-microdisk laser under mutually optical injection," *Appl. Phys. Lett.* **106**, 191107 (2015)
28. S. Ohara, A. K. Dal Bosco, K. Ugajin, A. Uchida, T. Harayama, and M. Inubushi, "Dynamics-dependent synchronization in on-chip coupled semiconductor lasers," *Phys. Rev. E* **96**, 032216 (2017)
29. M. Virte, K. Panajotov, H. Thienpont, and M. Sciamanna, "Deterministic polarization chaos from a laser diode," *Nat. Photonics* **7**, 60-65 (2013)
30. K. Kim, S. Bittner, Y. Zeng, S. Guazzotti, O. Hess, Q. J. Wang, and H. Cao, "Massively parallel ultrafast random bit generation with a chip-scale laser," *Science* **371**, 948-952 (2021)
31. C. G. Ma, J. L. Xiao, Z. X. Xiao, Y. D. Yang, Y. Z. Huang, "Chaotic microlasers caused by internal mode interaction for random number generation," *Light-Sci. Appl.* **11**, 187 (2022)
32. Y. Guo, Q. Cai, P. Li, R. Zhang, B. Xu, K. A. Shore, and Y. Wang, "Ultrafast and real-time physical random bit extraction with all-optical quantization," *Adv. Photonics* **4**, 035001 (2022).
33. A. Zhao, N. Jiang, J. Peng, S. Liu, Y. Zhang, and K. Qiu, "Parallel generation of low-correlation wideband complex chaotic signals using CW laser and external-cavity laser with self-phase-modulated injection," *Opto-Electron. Adv*. **5**, 200026 (2022).
34. E. Heidari, H. Dalir, M. Ahmed, V. J. Sorger, and R. T. Chen, "Hexagonal transverse-coupled-cavity VCSEL redefining the high-speed lasers," *Nanophotonics* **9**, 4743-4748 (2020)
35. H. Z. Weng, Y. Z. Huang, Y. D. Yang, X. W. Ma, J. L. Xiao, and Y. Du, "Mode Q factor and lasing spectrum controls for deformed square resonator microlasers with circular sides," *Phys. Rev. A* **95**, 013833 (2017).
36. H. Long, Y. Z. Huang, X. W. Ma, Y. D. Yang, J. L. Xiao, L. X. Zou, and B. W. Liu, "Dual-transverse-mode microsquare lasers with tunable wavelength interval," *Opt. Lett.* **40**, 3548-3551 (2015).
37. F. Y. Lin and J. M. Liu, "Nonlinear dynamical characteristics of an optically injected semiconductor laser subject to optoelectronic feedback," *Opt. Commun.* **221**, 173-180 (2003).
38. Y. L. Li, C. G. Ma, J. L. Xiao, T. Wang, J. L. Wu, Y. D. Yang, and Y. Z. Huang, "Wideband chaotic tri-mode microlasers with optical feedback," *Opt. Express* **30**, 2122-2130 (2022).
39. P. Grassberger, and I. Procaccia, "Characterization of Strange Attractors," *Phys. Rev. Lett.* **50**, 346-349 (1983).
40. K. Fraedrich and R. H. Wang, "Estimating the correlation dimension of an attractor from noisy and small datasets based on re-embedding," *Phys. D* **65**, 373-398 (1993).
41. NIST SP 800-22 statistical test suite, https://csrc.nist.gov/Projects/Random-Bit-Generation/Documentation-and-Software.
42. L. Li, A. Wang, P. Li, H. Xu, L. Wang, and Y. Wang, "Random bit Generator using delayed self-difference of filtered amplified spontaneous emission," *IEEE Photonics J.* **6**, 7500109 (2014).